\documentclass[amsmath,amssymb,11pt,aps,prd]{revtex4-2}
\usepackage{appendix}
\usepackage{booktabs}
\usepackage{dcolumn}
\usepackage{graphicx}
\usepackage[colorlinks,linkcolor=blue,anchorcolor=blue,urlcolor=blue,citecolor=blue]{hyperref}
\usepackage{makecell}
\usepackage{multirow}
\usepackage{orcidlink}
\usepackage{subfigure}
\usepackage{xcolor}
\usepackage{xfrac}

\begin{document}

\title{Research on gravitational waves from extreme mass ratio inspirals in an axisymmetric Finslerian extension of Schwarzschild spacetime}

\vspace{0.25in}
\author{Jin-Tao Yao\orcidlink{0009-0001-2013-3578}$^1$}
	\email{yaojintao@stu.cqu.edu.cn}

\author{Zhi-Chao Zhao\orcidlink{0000-0001-5180-4496}$^2$}
	\email{zhaozc@cau.edu.cn}

\author{Xin Li$^1$}
	\email[Corresponding author: ]{lixin1981@cqu.edu.cn}

\affiliation{$^1$ Department of Physics and Chongqing Key Laboratory for Strongly Coupled Physics, Chongqing University, Chongqing 401331, China	\\
$^2$ Department of Applied Physics, College of Science, China Agricultural University, 17 Qinghua East Road, Haidian District, Beijing 100083, China}

\vspace{0.25in}
\begin{abstract}

We investigate gravitational waves from extreme mass ratio inspirals in an axisymmetric Finslerian extension of the Schwarzschild spacetime. Its angular component is described by a two-dimensional Randers space, with the Schwarzschild limit recovered at Finslerian parameter $\epsilon = 0$. This structure breaks spherical symmetry, causing generic orbits to undergo orbital plane precession. Using multipolar radiation reaction and adiabatic orbital averaging, we derive the leading order Finslerian corrections to the energy, axial angular momentum, and total angular momentum fluxes. These corrections enter at Newtonian order and take different forms. We then incorporate these corrections into a higher-order Schwarzschild hybrid flux model to evolve the orbital elements and construct waveforms at mass quadrupole order using the numerical kludge method. The Finslerian corrections affect both the conservative orbital dynamics and the dissipative inspiral, producing cumulative phase shifts relative to the Schwarzschild waveform. For the representative source observed by LISA over one year with signal-to-noise ratio of 20, the noise weighted mismatch analysis shows that Finslerian waveforms with $|\epsilon| \gtrsim 2 \times 10^{-6}$ are distinguishable from their Schwarzschild counterparts, with higher initial eccentricities further lowering the distinguishable value of $|\epsilon|$. For a Schwarzschild injection, the Fisher matrix analysis yields a marginalized $1\sigma$ uncertainty of $\sigma_\epsilon \simeq 1.26 \times 10^{-6}$. These results show the potential of space-based gravitational wave detectors to probe the axisymmetric Finslerian extension of the Schwarzschild spacetime.

\end{abstract}

\maketitle

\section{Introduction}
Extreme mass ratio inspirals (EMRIs), in which a stellar-mass compact object (with mass $\mu$) spirals into a supermassive black hole (with mass $M$) with a mass ratio $\frac{\mu}{M}\lesssim 10^{-4}$, are among the most important sources for future space-based gravitational wave (GW) observatories, including the Laser Interferometer Space Antenna (LISA)~\cite{AmaroSeoane2018,LISA:2022yao}, TianQin~\cite{Luo2016,Li:2024rnk,Luo:2025ewp}, and Taiji~\cite{HUANG:2017vty,Hu2017}. During the inspiral, the secondary object completes $10^{4} \sim 10^{6}$ orbital cycles before plunging~\cite{Barack2004}. Over this long evolution, even small changes in the orbital frequencies and their secular drift can accumulate into observable modifications of the GW signal. The signal is therefore sensitive to both the conservative orbital dynamics determined by the background spacetime and the dissipative radiation reaction driving the inspiral, making EMRIs well suited for testing gravity in the strong-field regime~\cite{Cardenas2024}.

The use of EMRIs to probe the spacetime of a supermassive compact object was pioneered by Ryan, who showed that orbital observables and the GW spectrum encode the mass and current multipole moments of a generic stationary, axisymmetric source~\cite{Ryan1995Mapping}. Subsequent studies extended this program to quasi-Kerr, bumpy, and parametrized backgrounds~\cite{Glampedakis2006,Vigeland2011,Gair2011,xin2019}, as well as to generic deviations from the Kerr multipolar structure and other non-Kerr compact objects~\cite{Muguruza:2026zcs,Muguruza2026}. In these studies, the multipole moments characterizing the background geometry are distinct from the radiative multipoles used to generate the GW signal. In kludge models, the instantaneous orbital motion is described by geodesics of the background spacetime, whereas post-Newtonian (PN) or hybrid radiation reaction prescriptions govern the secular evolution of the orbital elements \cite{Glampedakis2002,Gair2006,Babak2007}. The waveform is then constructed from flat-spacetime radiative multipoles, typically at mass-quadrupole order~\cite{Glampedakis2006,Gair2011,xin2019,Fan2026,Liang2023,Fu2025,Zi2024jla,Yang2025esa,Muguruza2026}. Some implementations further incorporate Teukolsky-based fits to improve the accuracy of the radiation reaction fluxes~\cite{Canizares2012,Qiao2025}. In addition, black hole perturbation theory has been employed to compute both GW emission and radiation carried by additional fields, with the resulting fluxes incorporated into quadrupole, kludge, or fast adiabatic waveform models~\cite{Maselli2020,Maselli2021men,Guo2022,Barsanti2022,Barsanti2023,Zhang2023,Zi2024lmt,Speri2026,ZiLi2023}. More broadly, EMRI waveform modeling has advanced from augmented kludge constructions to fast relativistic and strong-field adiabatic models~\cite{Chua2017,Chua2021,Hughes2021,Katz2021}, alongside dedicated search and parameter estimation techniques for data analysis~\cite{Zhang2022,Ye2024}. Collectively, these developments provide a systematic theoretical and computational framework for using EMRIs to test the spacetime geometry of compact objects.

Most EMRI investigations have been conducted within general relativity (GR) or modified theories of gravity based on Riemannian geometry. As a natural generalization of Riemannian geometry, Finsler geometry provides a broader geometric framework for gravity by relaxing the quadratic restriction of the metric~\cite{bao2012introduction}. Since different formulations of Finsler gravity are generally inequivalent~\cite{Rutz1993,Pfeifer2012,Hohmann2019}, phenomenological studies must be carried out within a specific theoretical framework. In this work, we consider the exact axisymmetric Finslerian extension of Schwarzschild spacetime derived from the vacuum field equation proposed by Rutz~\cite{Li2014}, hereafter referred to as the Finslerian Schwarzschild spacetime.  Its angular component is described by a two-dimensional Randers space, and the Schwarzschild limit is recovered at Finslerian parameter $\epsilon=0$. This solution has been used to investigate quasinormal modes of scalar and electromagnetic test fields~\cite{LiZhao2020}, black hole shadows~\cite{He2024}, orbital dynamics \cite{Yao2023}, and the orbit of the S2 star around Sagittarius A$^*$~\cite{Li2022,Yao2025}. These studies have primarily focused on conservative dynamics and test-field propagation on a fixed background. To the best of our knowledge, radiation reaction driven EMRIs and their GW signals in the Finslerian Schwarzschild spacetime have not yet been investigated.

To address this gap, we develop a waveform model for radiation reaction driven EMRIs in the Finslerian Schwarzschild spacetime. Since a self-consistent formalism for source tensor perturbation and the associated GW fluxes has not been developed for this background, we adopt a hybrid treatment. We employ a two-timescale description that separates the rapid orbital motion from the slow radiation reaction driven evolution in the extreme mass ratio limit~\cite{Miller:2020bft}. On the orbital timescale, the instantaneous motion is governed by the Finslerian geodesic equations. On the radiation reaction timescale, the secular evolution of the semi-latus rectum, eccentricity, and inclination angle is obtained from the orbit averaged fluxes of the constants of motion. Specifically, we construct these fluxes by adding the leading order Finslerian corrections, derived using the multipolar radiation reaction formalism and adiabatic orbital averaging~\cite{Canizares2012,Qiao2025}, to the higher-order Schwarzschild hybrid fluxes of Ref.~\cite{Gair2006}. The resulting waveforms are generated at mass quadrupole order using the numerical kludge (NK) approach~\cite{Glampedakis2002,Babak2007}. In this construction, the Finslerian corrections are considered only in the source dynamics, while possible Finslerian corrections to GW propagation are neglected and GWs are assumed to propagate according to GR with only the two tensor polarizations. Following Ref.~\cite{He2024} and in view of the stringent local constraints on Lorentz violation~\cite{Will:2014kxa}, Lorentz-violating effects in the vicinity of Earth are considered negligible. Accordingly, the local spacetime surrounding the space-based GW detector is approximated as Minkowskian when modeling its response. Based on our previous studies of orbital dynamics in this spacetime~\cite{Li2022,Yao2023,Yao2025}, we predict that the effects of the Finslerian corrections will accumulate during the inspiral and leave potentially distinguishable imprints on EMRI waveforms. We model the detector output by applying the LISA response to the generated waveforms. We then assess the detectability of the Finslerian corrections using noise weighted mismatch to quantify waveform distinguishability and a Fisher matrix analysis to forecast constraints on the Finslerian parameter.

The remainder of this paper is organized as follows. In Sec.~\ref{sec.ii}, we review the Finslerian Schwarzschild spacetime and formulate its geodesics in terms of orbital elements and phase variables. In Sec.~\ref{sec.iii}, we derive the leading order Finslerian contributions to the orbit averaged fluxes and construct the adiabatic evolution equations. Section~\ref{sec.iv} presents the NK waveform, the LISA response, and the mismatch and Fisher matrix results. We summarize our conclusions and discuss the limitations of the present model in Sec.~\ref{sec.v}.

\section{Finslerian Schwarzschild Spacetime}\label{sec.ii}
Finsler geometry provides a natural generalization of Riemannian geometry by relaxing the requirement that the metric be quadratic in the tangent vectors~\cite{bao2012introduction}. The fundamental geometric quantity is the Finsler structure $F$, which satisfies the positive homogeneity condition $F(x,\lambda y)=\lambda F(x,y)$ for all $\lambda>0$, where $x$ denotes a point on the Finsler manifold $\mathcal{M}$ and $y\in T_x\mathcal{M}$ is a tangent vector. This property guarantees that the Finsler arc length $\int F(x(\tau),y(\tau)) d\tau$ is independent of the choice of curve parameter. The associated Finsler metric is defined by~\cite{bao2012introduction}
\begin{equation}
	g_{\mu\nu} \equiv \frac12 \frac{\partial^2 F^2}{\partial y^\mu \partial y^\nu} .
\end{equation}
Unlike Riemannian geometry, the Finsler structure is not required to be positive definite. Throughout this work we adopt the signature convention $(-+++)$, under which negative, zero, and positive values of $F^2$ correspond to timelike, null, and spacelike directions, respectively.

Several inequivalent gravitational field equations have been proposed in Finsler geometry~\cite{Rutz1993,Pfeifer2012,Hohmann2019}. In this work, we adopt the field equation proposed by Rutz~\cite{Rutz1993}, which is equivalent to requiring the Finsler Ricci scalar to vanish~\cite{Li2014},
\begin{equation}
	Ric = \frac{1}{F^2} \left( 2 \frac{\partial G^\mu}{\partial x^\mu} - y^\lambda \frac{\partial^2 G^\mu}{\partial x^\lambda \partial y^\mu} + 2 G^\lambda \frac{\partial^2 G^\mu}{\partial y^\lambda \partial y^\mu} - \frac{\partial G^\mu}{\partial y^\lambda} \frac{\partial G^\lambda}{\partial y^\mu}\right)	= 0		,
\end{equation}
where the geodesic spray coefficients $G^\mu$ are
\begin{equation}
	G^\mu = \frac{1}{4} g^{\mu \nu} \left( \frac{\partial^2 F^2}{\partial x^\lambda \partial y^\nu} y^\lambda - \frac{\partial F^2}{\partial x^\nu} \right).
\end{equation}
An exact vacuum solution of Rutz's field equation is the Finslerian Schwarzschild solution~\cite{Li2014},
\begin{equation}\label{Finsler Schwarzschild}
	F^2 = -f(r)y^ty^t + f(r)^{-1}y^ry^r + r^2\bar F^2	,
\end{equation}
where $f(r)=1-\frac{2GM}{r}$, and $M$ is the black hole mass. The angular component is described by a two-dimensional Randers Finsler space with constant positive flag curvature,
\begin{equation}\label{Finsler sphere}
	{\bar F} = \frac{\sqrt{(1-\epsilon^2\sin^2\theta)y^\theta y^\theta+\sin^2\theta y^\varphi y^\varphi}}{1-\epsilon^2\sin^2\theta}-\frac{\epsilon\sin^2\theta y^\varphi}{1-\epsilon^2\sin^2\theta}	,
\end{equation}
where the Finslerian parameter $|\epsilon|<1$ characterizes the deviation from the Riemannian two-sphere. In the limit $\epsilon \to 0$, the angular component reduces to the standard Riemannian two-sphere, and the spacetime recovers the Schwarzschild solution. It should be noted that two-dimensional Finsler space is irreversible under the transformation $y^\varphi \to -y^\varphi$. Therefore, positive and negative values of the Finslerian parameter $\epsilon$ in Eq.~\eqref{Finsler sphere} correspond to different geometrical structures~\cite{Li2022}.

The motion of a freely falling test particle is governed by the Finsler geodesic equation,
\begin{equation}
	\frac{d^2 x^\mu}{d\tau^2} + 2G^\mu = 0	.
\end{equation}
The geodesic motion admits three independent constants of motion, namely the specific energy $\mathcal E$, the specific axial angular momentum $\mathcal J$, and the total angular momentum $\mathcal K$, together with the normalization condition $F^2=-1$ for massive particles. The corresponding four-velocity components are
\begin{align}
	\frac{d t}{d\tau} &= \frac{\mathcal{E}}{f(r)}	,	\label{dt}\\
	\left(\frac{d r}{d\tau}\right)^2 &= \mathcal{E}^2 - f(r) \left(1 + \frac{\mathcal{K}^2}{r^2}\right)	,	\label{dr}\\
	\left(\frac{d \theta}{d\tau}\right)^2 &= \frac{\mathcal{K}^2}{r^4} \left( \frac{\sin^2\theta (\mathcal{K} - \epsilon \mathcal{J})^2 - \mathcal{J}^2}{\sin^2\theta (\mathcal{K} - \epsilon \mathcal{J})^2} \right) = \frac{\mathcal{K}^2}{r^4} \left( 1 - \frac{\cos^2\iota}{\sin^2\theta} \right)	,	\label{dtheta}\\
	\frac{d \varphi}{d\tau} &= \frac{\mathcal{K}}{r^2} \left( \frac{\epsilon \sin^2\theta (\mathcal{K} - \epsilon \mathcal{J}) + \mathcal{J}}{\sin^2\theta (\mathcal{K} - \epsilon \mathcal{J})} \right) = \frac{\mathcal{K}}{r^2} \left( \epsilon + \frac{\cos\iota}{\sin^2\theta} \right)	\label{dvarphi},
\end{align}
where $\iota = \frac{\pi}{2}-\theta_{\min}$ denotes the orbital inclination angle, and $\cos\iota = \frac{\mathcal J}{\mathcal K-\epsilon\mathcal J}$. The conserved quantities $\mathcal E$ and $\mathcal J$ arise from the Euler-Lagrange equations~\cite{shen2001lectures}, whereas $\mathcal K$ originates from the hidden symmetry of the Finslerian Schwarzschild spacetime.

The radial motion described by Eq.~\eqref{dr} possesses two turning points, namely the periastron $r_p$ and the apastron $r_a$, determined by $\frac{d r}{d\tau}=0$. Similarly, the polar motion is bounded by $\theta_{\min}$ and $\theta_{\max} = \pi-\theta_{\min}$, where the turning points satisfy $\frac{d\theta}{d\tau} = 0$. Following Ref.~\cite{Babak2007}, we introduce the monotonically increasing phase variables $\omega$ and $\chi$ to describe the radial and polar oscillations, respectively:
\begin{equation}\label{para_orbit}
	r = \frac{pM}{1 + e \cos\omega}	,	\qquad	\cos\theta = \sin\iota \sin\chi	,
\end{equation} 
where $p = \frac{2 r_a r_p}{M (r_a + r_p)}$ and $e = \frac{r_a - r_p}{r_a + r_p}$ denote the dimensionless semi-latus rectum and eccentricity, respectively. Substituting Eq.~\eqref{para_orbit} into Eqs.~\eqref{dt}-\eqref{dvarphi}, the geodesic equations become
\begin{align}
	\frac{d\omega}{d t} &= \frac{(p - 2 - 2 e \cos\omega) (1 + e \cos\omega)^2 \sqrt{p - 6 - 2 e \cos\omega}}{p^2 \sqrt{(p-2)^2 - 4 e^2}}	,	\label{domega} \\
	\frac{d\chi}{d t} &= \frac{(p - 2 - 2 e \cos\omega) (1 + e \cos\omega)^2}{p^{3/2} \sqrt{(p - 2)^2 - 4 e^2}}	,	\label{dchi} \\
	\frac{d\varphi}{d t} &= \frac{(p - 2 - 2 e \cos\omega) (1 + e \cos\omega)^2}{p^{3/2} \sqrt{(p-2)^2 - 4e^2}} \left(\epsilon + \frac{\cos\iota}{1 - (1 - \cos^2\iota) \sin^2\chi}\right)	,	\label{dvarphi2}
\end{align}
Within the adiabatic approximation, the orbital evolution naturally separates into two timescales. Eqs.~\eqref{domega}-\eqref{dvarphi2} govern the evolution of the phase variables $(\omega,\chi,\varphi)$, which describe the instantaneous geodesic motion; whereas gravitational radiation reaction drives the secular evolution of the orbital elements $(p,e,\iota)$. Stable bound orbits satisfy $p \ge 6+2e$, with equality defining the separatrix between stable inspiral and plunge.

\section{Fluxes and Adiabatic Evolution}\label{sec.iii}

Within the adiabatic approximation, the inspiral is modeled as a sequence of bound geodesics whose orbital constants evolve slowly under gravitational radiation reaction. The orbital period is assumed to be much shorter than the radiation reaction timescale, so that the secondary compact object follows an instantaneous geodesic characterized by a set of slowly varying orbital constants. The secular evolution of these constants is determined by the orbit averaged fluxes. 

Following the approaches developed in Refs.~\cite{Ryan1995,Ryan1996,Qiao2025}, we derive the evolution equations of the orbital constants from an effective Lagrangian description. The motion of a free particle is determined by the length of its worldline. Accordingly, the particle dynamics can be described by the proper time action
\begin{equation}
	S = -\mu \int d\tau	,
\end{equation}
where $\mu$ is the particle mass and $\tau$ is the proper time. Using the coordinate time $t$ as the evolution parameter, this action leads to the effective Lagrangian~\cite{Qiao2025}
\begin{equation}
	\mathcal{L} = \mu \left( 1 - \frac{d\tau}{dt} \right)	.
\end{equation}
Substituting the Finslerian Schwarzschild metric \eqref{Finsler Schwarzschild} into the effective Lagrangian and retaining terms through first PN order yields
\begin{equation}
	\mathcal{L} = \mu \left( \frac{1}{2} v_\epsilon^2 + \frac{M}{r}\right) + \mu \left(\frac{1}{8} v_\epsilon^4 + \frac{M}{2r} v_\epsilon^2 + \frac{M}{r} \dot{r}^2 + \frac{M^2}{2 r^2} \right) + \mathcal{O}(2PN)	,
\end{equation}
where $v_\epsilon^2=\dot r^{\,2}+r^2\bar F^2$ and an overdot denotes differentiation with respect to the coordinate time $t$.

Since the leading order Finslerian correction enters at Newtonian order, we retain terms linear in the Finslerian parameter $\epsilon$ and neglect mixed PN-Finslerian contributions. The effective Lagrangian therefore reduces to
\begin{equation}\label{eff_lag}
	\mathcal{L} = \frac{\mu}{2} \left( \dot{r}^2 + r^2 \dot{\theta}^2 + r^2 \sin^2\theta\dot{\varphi}^2 \right) + \frac{\mu M}{r} - \mu\epsilon r^2 \sin^2\theta\dot{\varphi} \sqrt{\dot{\theta}^2 + \sin^2\theta\dot{\varphi}^2}.
\end{equation}
The corresponding conserved energy and azimuthal angular momentum are obtained from the Euler-Lagrange equations,
\begin{align}
	E &= \dot{x}^i \frac{\partial \mathcal{L}}{\partial \dot{x}^i} - \mathcal{L} = \frac{\mu}{2} \left(  \dot{r}^2+r^2\dot{\theta}^2+r^2\sin^2\theta\dot{\varphi}^2 \right) - \frac{\mu M}{r} - \mu\epsilon r^2\sin^2\theta\dot{\varphi} \sqrt{\dot{\theta}^2 + \sin^2\theta\dot{\varphi}^2}	,	\\
	J &= \frac{\partial \mathcal{L}}{\partial \dot{\varphi}} = \mu r^2 \sin^2 \theta \dot{\varphi} - \mu \epsilon r^2 \sin^2\theta \frac{\dot{\theta}^2 + 2\sin^2\theta\dot{\varphi}^2}{\sqrt{\dot{\theta}^2 + \sin^2\theta\dot{\varphi}^2}}	.
\end{align}
The third constant of motion follows from the Hamilton-Jacobi equation and is given by
\begin{align}
	K = \mu r^2 \left( \sqrt{\dot{\theta}^2 + \sin^2\theta\dot{\varphi}^2} - \epsilon \sin^2\theta \dot{\varphi} \right)	.
\end{align}
The conserved quantities $\{ E,J,K \}$ are the coordinate time counterparts of the $\{ \mathcal{E}, \mathcal{J}, \mathcal{K} \}$, since both formulations originate from the Finslerian particle action~\cite{Ryan1995,misner1973gravitation}. At the Newtonian order used to derive the leading order Finslerian flux corrections, the effective constants satisfy
\begin{equation}
	E / \mu \simeq \mathcal{E} - 1	,	\qquad	J / \mu \simeq \mathcal{J} / u^t	,	\qquad	K / \mu \simeq \mathcal{K} / u^t	,
\end{equation}
where $u^t = dt/d\tau$ is given by Eq.~\eqref{dt}. At the lowest PN order, $u^t \approx 1 + \mathcal{O}(PN)$, so that $J / \mu = \mathcal{J}$ and $K / \mu = \mathcal{K}$ to the order considered in this work. Consequently, the orbit averaged fluxes derived below can be consistently combined with the geodesic equations in Sec.~\ref{sec.ii} to describe the adiabatic evolution of EMRIs.

Following Ref.~\cite{Ryan1996}, we introduce the auxiliary Cartesian coordinates to derive the radiative evolution of the orbital constants,
\begin{equation}\label{auxiliary}
	x^i = r \left( \sin\theta\cos\varphi,\, \sin\theta\sin\varphi,\, \cos\theta \right)	.
\end{equation}
For convenience, we define $h_i = \varepsilon_{ijk} x_j \dot{x}_k$, $h = \sqrt{h_i h_i}$, and $h_z = h_3$, with $r^2 = x_i x_i$. Here, $\varepsilon_{ijk}$ is the Levi-Civita tensor, repeated indices indicate summation, and an overdot denotes differentiation with respect to coordinate time. Differentiating the orbital constants with respect to coordinate time gives
\begin{align}
	\dot{E} &= \mu v_i a_i + \frac{\mu M x_i v_i}{r^3} - \mu \epsilon \left[ \frac{h_z h_i}{h r^2} \varepsilon_{ijk} x_j a_k + \frac{h}{r^2} \varepsilon_{3jk} x_j a_k - \frac{2h h_z}{r^4} x_i v_i \right]	,	\label{eff_dE}	\\
	\dot{J}	&= \mu \varepsilon_{3jk} x_j a_k - \mu \epsilon \left[ \left( \sin^2\theta - \frac{h_z^2}{h^2} \right) \frac{h_i}{h} \varepsilon_{ijk} x_j a_k + \frac{2 h_z}{h} \varepsilon_{3jk} x_j a_k  + h \frac{d \sin^2\theta}{dt}\right]	,	\label{eff_dJ}	\\
	\dot{K} &= \mu \frac{h_i}{h} \varepsilon_{ijk} x_j a_k - \mu \epsilon \varepsilon_{3jk} x_j a_k	\label{eff_dK}	,
\end{align}
where $v_i = \dot{x}_i$ is the velocity and $a_i = \ddot{x}_i$ is the acceleration. The acceleration can be decomposed into conservative and radiation reaction contributions,
\begin{equation}
	a_i = a^{\rm cons}_i + a^{\rm RR}_i	.
\end{equation}
The conservative dynamics do not contribute to the secular evolution of the orbital constants. The contributions associated with GW emission are governed by $a^{\rm RR}_i$, which can be obtained by making use of the Burke-Thorne generalized force~\cite{Thorne:1969rba,Burke:1970wx,Galley2009,Galley2012}
\begin{equation}
	{\mathcal{F}}^{\rm RR}_i = -\frac{2\mu}{5} Q^{(5)}_{ij} \boldsymbol{x}_j	,
\end{equation}
where $Q_{ij} = I_{ij} - \mu \delta_{ij} r^2 / 3$ is the symmetric trace-free part of the mass quadrupole moment $I_{ij} = \mu x_i x_j$. The forced Euler-Lagrange equation then yield
\begin{equation}
	a^{\rm RR}_i = \delta_{ik} \left(\mathcal{H}^{-1}\right)^{kj} {\mathcal{F}}^{\rm RR}_j	,
\end{equation}
where $\delta_{ik}$ is the Kronecker delta and $\mathcal{H}_{ij} = \frac{\partial^2 \mathcal{L}}{\partial v_i \partial v_j}$ is the velocity Hessian. The detailed derivation is provided in Appendix~\ref{appendix.i}.

Within the adiabatic approximation, we obtain the GW fluxes by averaging Eqs.~\eqref{eff_dE2}-\eqref{eff_dK2}, derived from Eqs.~\eqref{eff_dE}-\eqref{eff_dK}, over the orbital period using the orbital parameterization \eqref{para_orbit}. This yields the Newtonian order fluxes together with the corresponding leading order Finslerian corrections. Although these corrections capture the dominant dissipative effect of the Finslerian parameter, they do not include the higher-order PN contributions required for accurate EMRI waveform modeling. Since a complete treatment of GW emission in the Finslerian Schwarzschild spacetime is not yet available, we incorporate the Finslerian corrections into the hybrid flux prescription proposed by Glampedakis \textit{et al.}~\cite{Glampedakis2002} and subsequently refined by Gair and Glampedakis~\cite{Gair2006}. In this prescription, the higher-order PN contributions are taken from the hybrid model, while the leading order Finslerian corrections are added to the corresponding fluxes. The resulting hybrid fluxes are
\begin{align}
	\langle \frac{dE}{dt} \rangle = &- \frac{32 \mu^2}{5 M^2} \frac{(1-e^2)^{3/2}}{p^5} \left[ g_1(e) - \frac{1}{p} g_3(e) + \frac{\pi}{p^{3/2}} g_4(e) - \frac{1}{p^2} g_5(e) + \frac{1}{p^{5/2}}\left( c_1^a + \frac{c_1^b}{p^{1/2}} + \frac{c_1^c}{p} \right) \right]	\nonumber	\\
	&- \frac{8 \mu^2}{5 M^2} \frac{\epsilon \cos\iota (1-e^2)^{3/2}}{p^5} \left( 24 + 104 e^2 + 33 e^4 \right),	\label{eff_average_dE} \\
	\langle \frac{dJ}{dt} \rangle = &- \frac{32 \mu^2}{5 M} \frac{\cos\iota (1-e^2)^{3/2}}{p^{7/2}} \left[ g_9(e) - \frac{1}{p} g_{11}(e) + \frac{\pi}{p^{3/2}} g_{12}(e) - \frac{1}{p^2} g_{13}(e) + \frac{1}{p^{5/2}}\left( c_1^a + \frac{c_1^b}{p^{1/2}} + \frac{c_1^c}{p} \right) \right]	\nonumber	\\
	&- \frac{2 \mu^2}{5 M} \frac{\epsilon (1-e^2)^{3/2}}{p^{7/2}} \left[ 20 + 33 e^2 + 2 e^4 + \cos^2\iota \left(60 + 93 e^2 + 2 e^4\right)  \right],	\label{eff_average_dJ} \\
	\langle \frac{dK}{dt} \rangle = &- \frac{32 \mu^2}{5 M} \frac{(1-e^2)^{3/2}}{p^{7/2}} \left[ g_9(e) - \frac{1}{p} g_{11}(e) + \frac{\pi}{p^{3/2}} g_{12}(e) - \frac{1}{p^2} g_{13}(e) + \frac{1}{p^{5/2}}\left( c_1^a + \frac{c_1^b}{p^{1/2}} + \frac{c_1^c}{p} \right) \right]	\nonumber	\\
	&- \frac{8 \mu^2}{5M} \frac{\epsilon \cos\iota (1-e^2)^{3/2}}{p^{7/2}} \left(24 + 35 e^2 + e^4\right),	\label{eff_average_dK}
\end{align}
where the functions $g_i(e)$ and the coefficients $c_1^i$ are given in Refs.~\cite{Gair2006,Canizares2012}. The coefficients $c_1^i$ were obtained by fitting to numerical solutions of the Teukolsky equation, thereby reproducing the correct behavior in the near-circular and near-polar limits.

\section{Waveform Modeling and Parameter Estimation}\label{sec.iv}

The orbit averaged fluxes derived in Sec.~\ref{sec.iii} govern the secular evolution of the orbital elements $p(t)$, $e(t)$, and $\iota(t)$. Combined with the geodesic equations for the phase variables $\omega(t)$, $\chi(t)$, and $\varphi(t)$, Eqs.~\eqref{domega}-\eqref{dvarphi2}, they determine the adiabatic trajectory of the secondary compact object. Based on this trajectory, we construct the corresponding gravitational waveform and investigate its observational consequences.

A self-consistent perturbative description of GW generation in the Finslerian Schwarzschild spacetime is not yet available. We therefore adopt the NK prescription developed in Ref.~\cite{Babak2007}, in which the orbital trajectory is mapped onto spherical polar coordinates in an auxiliary flat spacetime and the emitted radiation is computed from the corresponding multipole moments. We further assume that the GW signals, including the leading order Finslerian corrections, propagate as the two tensor polarizations of GR. Within this framework, the Finslerian correction affects the waveform through both the conservative orbital dynamics and the dissipative GW fluxes.

Using the auxiliary Cartesian coordinates and mass quadrupole moment introduced in Sec.~\ref{sec.iii}, we compute the spatial components of the trace-reversed metric perturbation at quadrupole order
\begin{equation}
	\bar{h}_{ij}(t) = \frac{2}{D_L} \ddot{I}_{ij}(t') \Big|_{t'=t-D_L},
\end{equation}
where $D_L$ is the luminosity distance and overdots denote differentiation with respect to the coordinate time $t$. Following the polarization convention adopted in the EMRI Kludge Suite~\cite{Babak2007,EKS}, we obtain the two GW polarization by projecting $\bar{h}_{ij}$ onto the polarization basis:
\begin{align}
	h_+ =&{\,} \frac{1}{2} \Bigl[ - \cos^2\lambda \left( \bar{h}_{xx} \sin^2\phi_0 + \bar{h}_{xy} \sin2\phi_0 + \bar{h}_{yy} \cos^2\phi_0 \right) - \bar{h}_{zz} \sin^2\lambda	\nonumber	\\
	&{\,} +\sin2\lambda \left( \bar{h}_{xz} \sin\phi_0 + \bar{h}_{yz} \cos\phi_0 \right) + \left( \bar{h}_{xx} \cos^2\phi_0 - \bar{h}_{xy} \sin2\phi_0 + \bar{h}_{yy} \sin^2\phi_0 \right) \Bigr],	\\ 
	h_\times =&{\,} -\frac{1}{2} \cos\lambda \left( \bar{h}_{xx} \sin2\phi_0 + 2 \bar{h}_{xy} \cos2\phi_0 - \bar{h}_{yy} \sin2\phi_0 \right) + \sin\lambda \left( \bar{h}_{xz} \cos\phi_0 - \bar{h}_{yz} \sin\phi_0 \right)	.
\end{align}
The viewing angle $\lambda$ is the angle between the source direction and the symmetry axis of the central black hole, which are specified in the Solar System barycentric frame by $(\theta_S, \phi_S)$ and $(\theta_K, \phi_K)$, respectively. It is given by $\cos\lambda = \cos\theta_S \cos\theta_K + \sin\theta_S \sin\theta_K \cos(\phi_S - \phi_K)$. The angle $\phi_0$ characterizes the orbit's initial azimuthal orientation about the black hole's symmetry axis relative to the line of sight.

\begin{figure*}[htbp]
	\centering
	\includegraphics[width=0.95\linewidth]{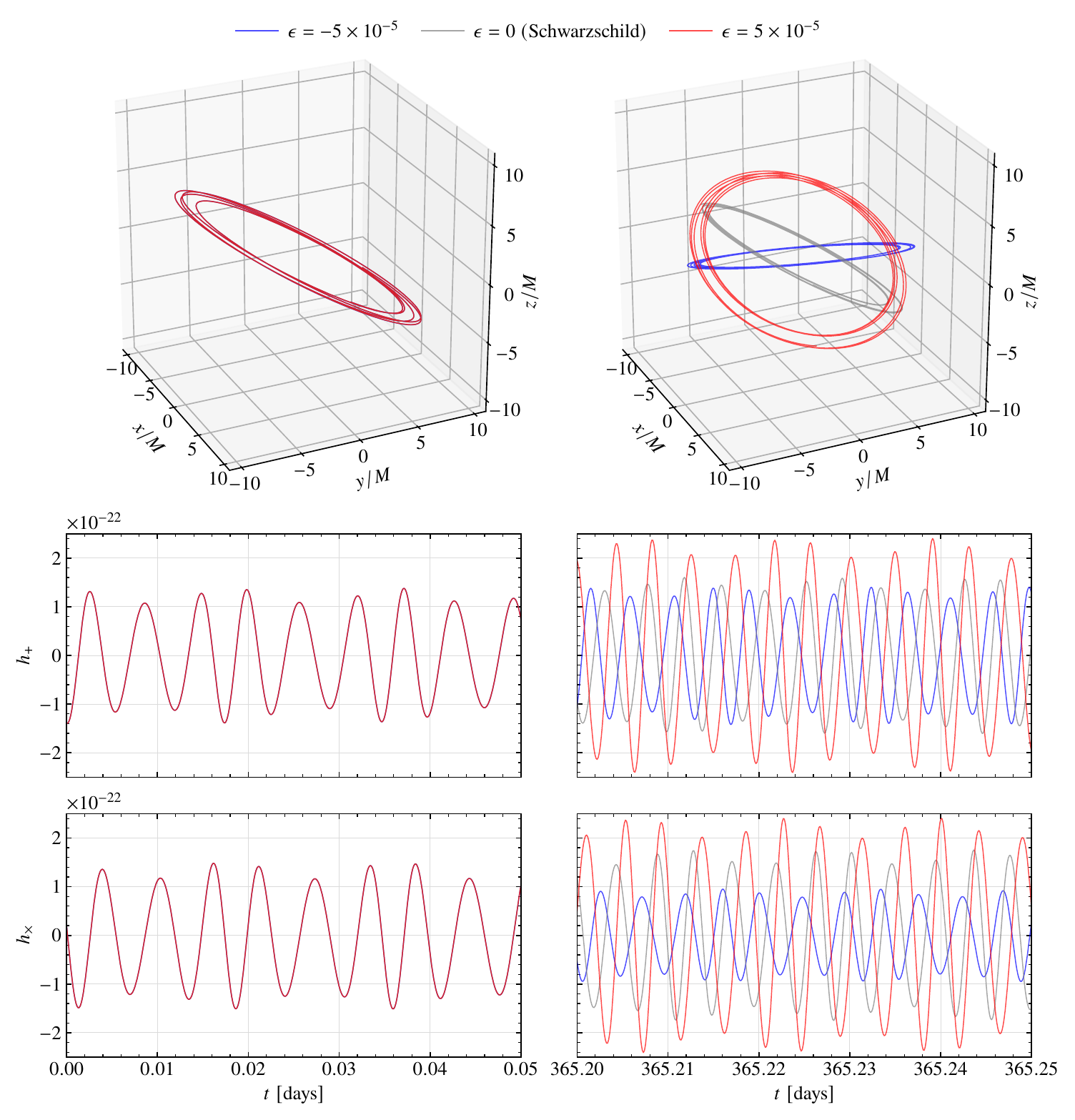}
	\caption{3D orbital trajectories (top panels) and the corresponding GW polarizations $h_+$ (middle panels) and $h_\times$ (bottom panels) for $\epsilon = 0$ (Schwarzschild) and $\epsilon = \pm 5 \times 10^{-5}$. The left and right columns show segments at the beginning and end of the one year evolution, respectively. The trajectories and polarizations initially nearly coincide, whereas accumulated Finslerian effects produce clear differences in trajectories and waveform at late times.}
	\label{WF_examples}
\end{figure*}

Figure~\ref{WF_examples} shows the secondary black hole's trajectories and the corresponding GW polarizations for $\epsilon = 0$, $\pm 5 \times 10^{-5}$. We set $M=10^6M_\odot$, $\mu=10M_\odot$, and $D_L=1~{\rm Gpc}$, with the source position and the orientation of the central black hole symmetry axis fixed to $(\theta_S,\phi_S) = \left( \pi / 3, \pi / 2 \right)$ and $(\theta_K,\phi_K) = \left( \pi / 4, \pi / 4 \right)$, respectively. The initial orbital elements are chosen as $(p_0,e_0,\iota_0) = \left( 10, 0.1, \pi / 6 \right)$, and the evolution spans one year. The trajectories and polarizations are indistinguishable at the beginning of the evolution, but accumulated Finslerian corrections produce clear differences in orbital orientation and waveform phase at late times. For the same initial configuration, opposite signs of $\epsilon$ produce distinct trajectories and polarizations. This sign dependence reflects the nonreversibility of the Randers space under reversal of the azimuthal direction, reflecting the directional dependence of the Randers space. These results indicate that long duration observations of EMRI GWs provide a promising opportunity to distinguish inspirals associated with different values of the Finslerian parameter.

\begin{figure*}[htbp]
	\centering
	\includegraphics[width=0.95\linewidth]{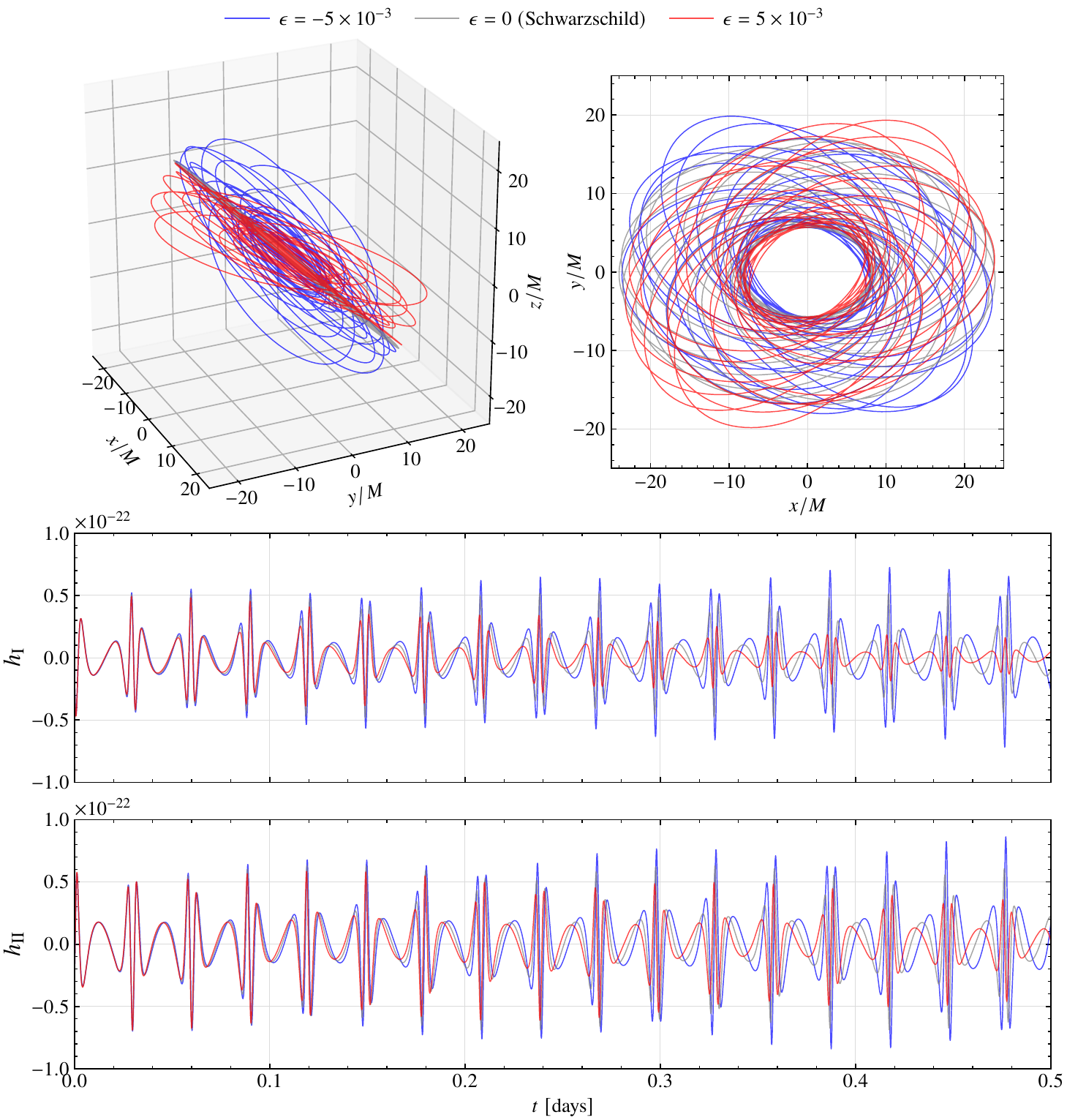}
	\caption{3D trajectories (upper left), their projections onto the $x-y$ plane (upper right), and the LISA Michelson-like channel signals $h_{\rm I}$ (middle) and $h_{\rm II}$ (bottom) for $\epsilon = 0$ (Schwarzschild), $\pm 5 \times 10^{-3}$. The panels show the first 0.5 days of the one year evolution. The Schwarzschild trajectory remains in a fixed orbital plane, while the Finslerian trajectories exhibit orbital plane precession and produce distinct responses in detector channels.}
	\label{WF_orbit_LISA}
\end{figure*}

To assess the observational consequences, we incorporate the detector response into the waveform model. The detector response is evaluated using the LISA response function, yielding two independent Michelson-like interferometer channels~\cite{Cutler1998,CornishRubbo2003},
\begin{equation}\label{lisa_response}
	h_\beta(t)=\frac{\sqrt{3}}{2}\left[F^+_\beta(t)h_+(t)+F^\times_\beta(t)h_\times(t)\right],	\qquad \beta=I,II.
\end{equation}
where $F^+_\beta(t)$ and $F^\times_\beta(t)$ are the antenna pattern functions, whose explicit expressions are given in Ref.~\cite{Barack2004}. Figure~\ref{WF_orbit_LISA} compares the secondary black hole's trajectories and corresponding LISA channel signals for $\epsilon = 0$, $\pm 5 \times 10^{-3}$. We set $(\theta_K,\phi_K) = \left( \pi / 4, \pi / 5 \right)$ and $(p_0,e_0,\iota_0) = \left( 12, 0.5, \pi / 4 \right)$, with all other parameters are the same as in Figure~\ref{WF_examples}. The 3D trajectories and their projection onto the $x-y$ plane are shown alongside the two Michelson-like channel responses over the first 0.5 days of evolution. The Schwarzschild trajectory remains confined to a fixed orbital plane, whereas the Finslerian correction changes the orientation of the instantaneous orbital plane over time. Together with the accumulated orbital phase difference, this change produces visible differences in both LISA channel signals.

The detector response constructed above enables a quantitative assessment of the observational consequences of the Finslerian correction. We first investigate the distinguishability between waveforms using the mismatch function. For a simulated signal $h_S(t)$ and a template waveform $h_T(t)$, their noise weighted inner product is defined as~\cite{Cutler1994,Balasubramanian1996}
\begin{equation}\label{inner_product}
	(h_S,h_T) = 2\int_0^\infty \frac{\tilde h_S^{\,*}\tilde h_T+\tilde h_S\tilde h_T^{\,*}}{S_n(f)} df	,
\end{equation}
where tildes denote Fourier transforms and $S_n(f)$ is the one-sided LISA noise power spectral density~\cite{Robson2019}. To remove arbitrary relative offsets in time and phase, we introduce a time shift $t_s$ and a phase shift $\phi_s$, such that 
\begin{equation}
	\tilde{h}^{(t_s, \phi_s)}_T = \tilde{h}_T e^{i\left(2\pi f t_s + \phi_s\right)}	.
\end{equation}
The normalized overlap is then defined as
\begin{equation}
	\mathcal{O}(t_s,\phi_s) = \frac{(h_S,\tilde{h}^{(t_s, \phi_s)}_T)}{\sqrt{(h_S,h_S)(h_T,h_T)}}	.
\end{equation}
The match is obtained by maximizing the overlap over the relative time and phase shifts~\cite{Read2009,Datta2024}
\begin{equation}
	\mathcal{O}_{\max} = \max_{t_s,\phi_s} \mathcal{O}(t_s,\phi_s)	, 
\end{equation}
and the corresponding mismatch is 
\begin{equation}
	\mathcal{M} = 1 - \mathcal{O}_{\max}	.
\end{equation}
This maximization gives $0 \leqslant \mathcal{O}_{\max} \leqslant 1$ and hence $0 \leqslant \mathcal{M} \leqslant 1$, where $\mathcal{M} = 0$ corresponds to waveforms that are identical up to relative time and phase shifts~\cite{Barausse2007}.

\begin{figure*}[htbp]
	\centering
	\includegraphics[width=0.49\linewidth]{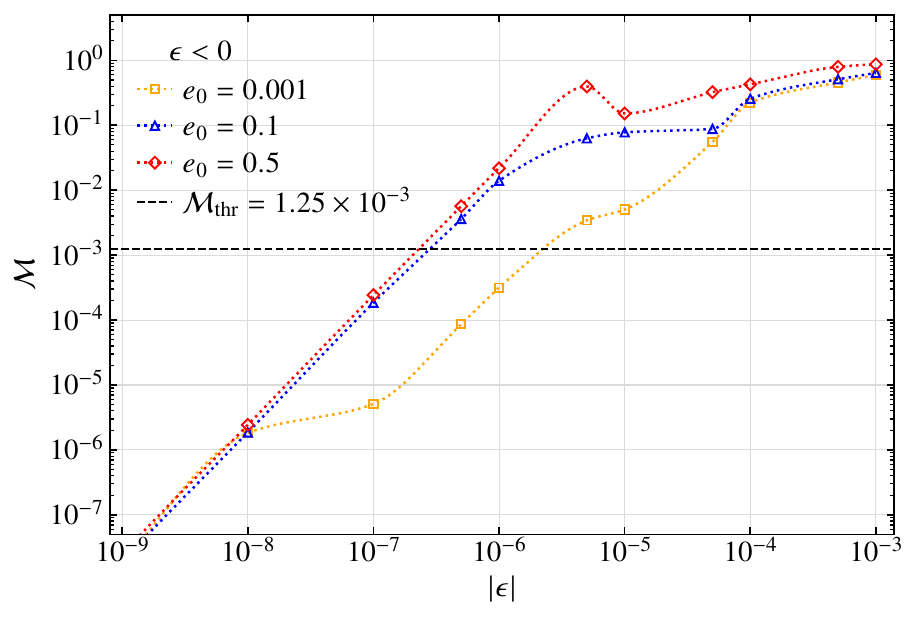}
	\includegraphics[width=0.49\linewidth]{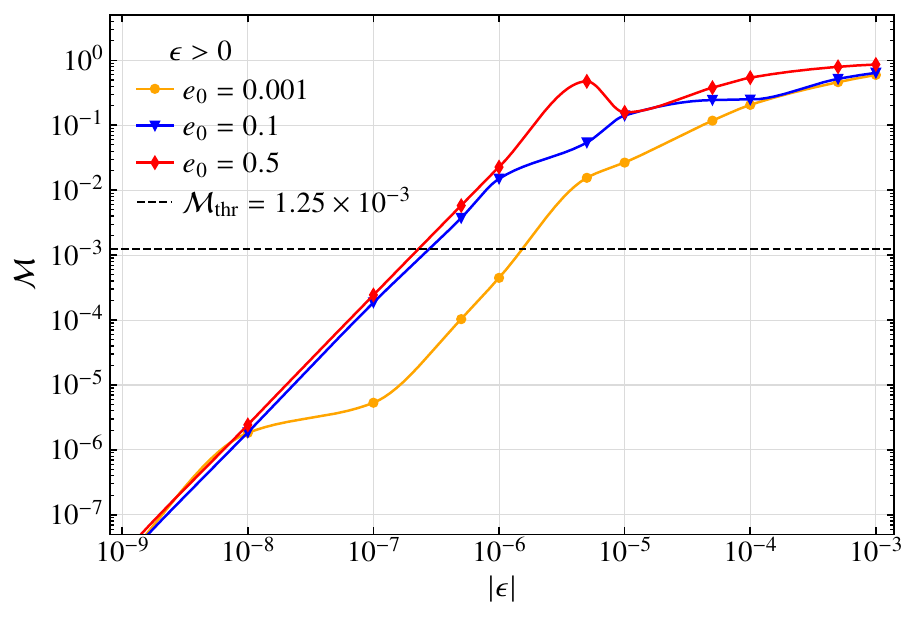}
	\caption{One year noise weighted mismatch between the Finslerian and Schwarzschild waveforms as a function of the Finslerian parameter $\epsilon$ for three initial eccentricities. The left and right panels correspond to negative and positive $\epsilon$, respectively. The horizontal black dashed line marks the mismatch threshold $\mathcal{M}_{\rm thr} = 1.25 \times 10^{-3}$ for a signal with an SNR=20 accumulated over one year observation.}
	\label{mismatch_epsilon}
\end{figure*}

Following Refs.~\cite{Maselli2021men,Zi2023}, two waveforms can be regarded as distinguishable by LISA when their mismatch satisfies $\mathcal{M} \gtrsim \mathcal{M}_{\rm thr} = \frac{D/2}{\rm SNR^2}$, where $D$ is the effective number of parameters allowed to vary in the comparison. Adopting SNR=20 and $D=1$ gives the one-parameter distinguishability mismatch threshold $\mathcal{M}_{\rm thr}=1.25\times10^{-3}$. For both the $\epsilon$- and time-dependent analyses, we fix $M=10^6M_\odot$, $\mu=10M_\odot$, $p_0=10$, and $\iota_0=\pi/6$. Figure~\ref{mismatch_epsilon} shows the one year mismatch between the Finslerian and Schwarzschild waveforms as a function of Finslerian parameter $\epsilon$ for $e_0 \in \{ 0.001, 0.1, 0.5 \}$, with negative and positive $\epsilon$ shown in the left and right panels, respectively. The mismatch lies well below the threshold for sufficiently small $|\epsilon|$ and generally increases with $|\epsilon|$, although local decreases occur at larger values. These decreases reflect the nonmonotonic dependence of the orbital deviation on $\epsilon$, similar to the behavior reported for stellar orbital precession~\cite{Li2022}. Positive and negative values of $\epsilon$ produce quantitatively different mismatches, while larger initial eccentricities generally yield greater distinguishability for the configurations considered. Despite the local decreases, the mismatch remains above $\mathcal{M}_{\rm thr}$ for all three eccentricities when $|\epsilon| \gtrsim 2 \times 10^{-6}$ over the range shown. These results indicate that long duration observations of EMRI GWs can distinguish the corresponding Finslerian waveforms from their Schwarzschild counterparts.

\begin{figure*}[htbp]
	\centering
	\includegraphics[width=0.48\linewidth]{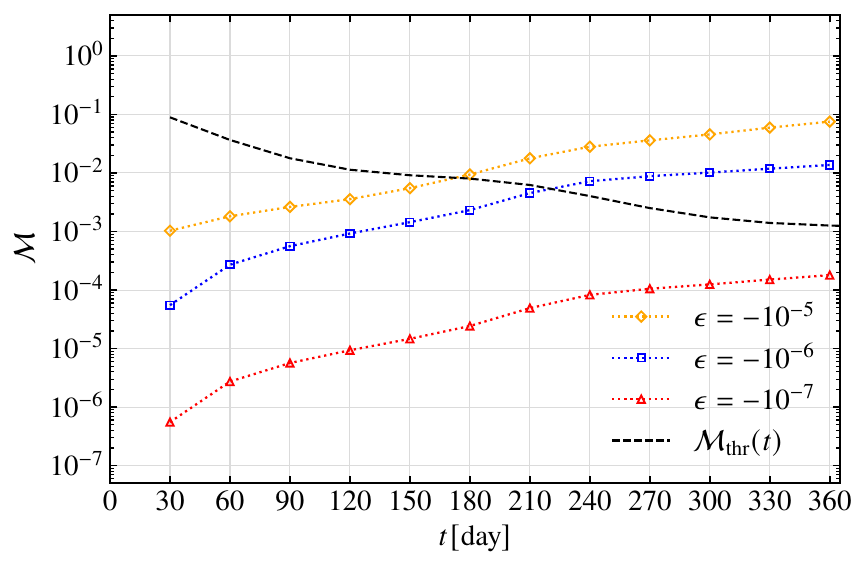}
	\includegraphics[width=0.48\linewidth]{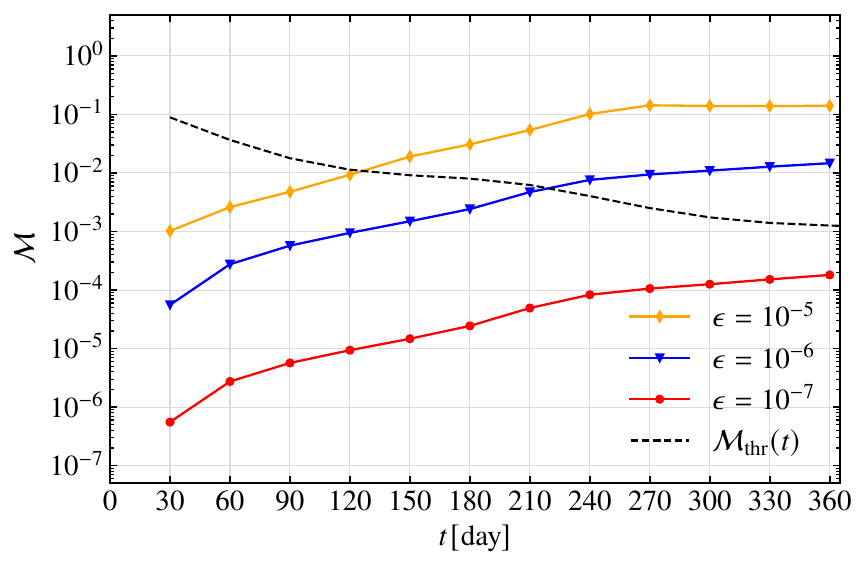}
	\caption{Cumulative noise weighted mismatch $\mathcal{M}$ between the Finslerian and Schwarzschild waveforms as a function of observation time for negative (left panel) and positive (right panel) values of $\epsilon$, with $e_0=0.1$. The black dashed lines mark the mismatch threshold $\mathcal{M}_{\rm thr}(t)$ computed from the SNR accumulated at each observation time. }
	\label{mismatch_time}
\end{figure*}

Figure~\ref{mismatch_time} presents the evolution of the mismatch between Finslerian and Schwarzschild waveforms with observation time for the cases in Figure~\ref{mismatch_epsilon} with $e_0 = 0.1$ and $\epsilon = \left\{\pm 10^{-7}, \pm 10^{-6}, \pm 10^{-5}\right\}$. The mismatch generally increases with observation time as waveform differences accumulate, while the corresponding distinguishability threshold decreases. The threshold is crossed after approximately $180$ days for $\epsilon=-10^{-5}$ and $130$ days for $\epsilon=10^{-5}$. For $|\epsilon|=10^{-6}$, the crossing occurs after approximately $220$ days for both signs of $\epsilon$. In contrast, the mismatch for $|\epsilon|=10^{-7}$ remains below the threshold throughout the one year observation, consistent with Figure~\ref{mismatch_epsilon}. These results indicate that longer observations improve waveform distinguishability, showing the potential of EMRI GW signals to probe smaller Finslerian corrections.

The mismatch analysis assesses differences between Finslerian and Schwarzschild waveforms after maximizing the overlap over relative time and phase offsets, with all remaining parameters held fixed. It neither accounts for differences that could be absorbed by varying other source parameters nor directly estimates the measurement uncertainty in $\epsilon$. We therefore complement this comparison with a Fisher matrix analysis to estimate statistical uncertainties of $\epsilon$. In the high SNR limit, the inverse Fisher matrix provides a local approximation to the parameter covariance matrix, yielding estimates of statistical uncertainties and parameter correlations~\cite{Barack2004}. For a waveform $h$ described by the parameter vector $\boldsymbol{\lambda}$, the Fisher matrix is defined as
\begin{equation}
	\Gamma_{ij} = \left( \frac{\partial h}{\partial\lambda_i}, \frac{\partial h}{\partial\lambda_j} \right)	,
\end{equation}
where the noise weighted inner product is given by Eq.~\eqref{inner_product}. Assuming stationary Gaussian detector noise, we estimate the marginalized $1\sigma$ measurement uncertainties using the Cramér-Rao bound as $\sigma_{\lambda_i}=\sqrt{(\Gamma^{-1})_{ii}}$~\cite{Vallisneri2008}. In the present analysis, we take the Fisher parameter vector to be $\boldsymbol{\lambda}=(M,\mu,p_0,e_0,\epsilon)$ and fiducial values $\boldsymbol{\lambda}_{\rm fid} = \left(10^6 M_\odot, 10 M_\odot, 10, 0.1, 0\right)$ corresponding to a Schwarzschild injection. All remaining waveform parameters are held fixed, with $\iota_0=\pi/6$, $(\theta_S,\phi_S)= (\pi/3,\pi/2)$, and $(\theta_K,\phi_K)=(\pi/4,\pi/4)$. The fiducial luminosity distance is determined by requiring the signal to have an $\mathrm{SNR} = 20$ over one year LISA observation. The resulting marginalized $1\sigma$ uncertainty of the Finslerian parameter is $\sigma_\epsilon \simeq 1.26 \times 10^{-6}$ at $\mathrm{SNR} = 20$. This forecast approximately two orders of magnitude tighter than the $\mathcal{O}(10^{-4})$ constraints from Galactic Center stellar orbit observations~\cite{Li2022,Yao2025}, although the analyses use different data and modeling assumptions. Together with the mismatch analysis, these findings suggest that the long duration EMRI observations have the potential to constrain the Finslerian Schwarzschild spacetime.

\section{Conclusion and Discussion}\label{sec.v}
In this work, we investigated GW signals from EMRIs in the Finslerian Schwarzschild spacetime. The two-dimensional Randers space breaks spherical symmetry, leading to orbital plane precession for generic orbits~\cite{Li2022,Yao2023,Yao2025}. Within a two-timescale description, the instantaneous orbital motion is governed by the Finslerian geodesic equations, whereas the secular evolution of the orbital elements is driven by orbit averaged fluxes. We derived the leading order Finslerian corrections to energy, axial angular momentum, and total angular momentum fluxes using the multipolar radiation reaction formalism and adiabatic orbital averaging. These corrections enter at Newtonian order and were incorporated into a higher-order Schwarzschild hybrid fluxes. The resulting inspiral trajectories were then used to construct mass quadrupole order waveforms with the NK approach and to compute the corresponding LISA response.

The Finslerian corrections modify both the conservative orbital dynamics and the dissipative inspiral, leading to accumulated phase differences relative to Schwarzschild waveforms. For the representative one year LISA observations with SNR=20, the mismatch analysis indicates that Finslerian waveforms with $|\epsilon| \gtrsim 2 \times 10^{-6}$ are distinguishable from their Schwarzschild counterparts across the eccentricities considered, with sensitivity extending to $|\epsilon| \gtrsim 2 \times 10^{-7}$ for the systems with $e_0 \sim \mathcal{O}(10^{-1})$. For $e_0 = 0.1$, the time-dependent analysis shows that approximately 220 days of observation suffice to distinguish waveforms with $|\epsilon| = 10^{-6}$, whereas those with $|\epsilon| = 10^{-4}$ require fewer than $90$ days. Complementing these comparisons, the Fisher matrix analysis of an injected Schwarzschild signal yields a marginalized $1 \sigma$ uncertainty of $\sigma_\epsilon \simeq 1.26 \times 10^{-6}$. By comparison, nearly three decades of observations of the S2 stellar orbit around the Galactic Center constrain the Finslerian parameter at the $\mathcal{O}(10^{-4})$ level~\cite{Li2022,Yao2025}. The EMRI results thus suggest both a substantially shorter observation time for distinguishing corrections at this level and a potential improvement of approximately two orders of magnitude in parameter precision over a one year observation. Although waveform distinguishability and parameter constraints are different measures, and the analyses rely on different modeling assumptions, these complementary results show the potential of EMRI observations to probe Finslerian corrections.

One improvement to the present framework is to develop a self-consistent tensor perturbation formalism for the Finslerian Schwarzschild spacetime. The mean metric of a Finsler space introduced in Ref.~\cite{HE20171162} provide a basis for constructing a tensor perturbation equations, extending the existing analyses of scalar and electromagnetic perturbations~\cite{LiZhao2020,Li2022QNM}. Such a formalism would enable direct calculations of GW fluxes and a systematic study of Finslerian effects on GW generation and propagation, supporting the development of more accurate relativistic waveforms. These waveforms could then be used in Bayesian analyses of a wider range of EMRI systems to assess parameter degeneracies and obtain more robust constraints on the Finslerian parameter.

\begin{acknowledgments}
	This work has been supported by the National Natural Science Foundation of China (Grants No. 12275034 and No. 12547101). One of the authors Jin-Tao Yao thanks Dr. Tian-Yang Li for helpful discussions.
\end{acknowledgments}

\appendix
\section{Finslerian corrections to the radiation reaction and fluxes}\label{appendix.i}
In this appendix, we derive the leading-order Finslerian correction to the radiation reaction acceleration and obtain the corresponding instantaneous rates of change of the orbital constants $\left\{E,J,K\right\}$. These rates provide the input for the orbital-averaging procedure used to calculate the secular fluxes.

Within the adiabatic approximation, changes in the orbital elements $\{p,e,\iota\}$ can be neglected over the orbital period. We therefore compare Finslerian and Schwarzschild geodesics with identical orbital elements and initial phases $\{\omega_0,\chi_0,\varphi_0\}$. Unlabelled quantities refer to the Finslerian orbit, while the subscript ``S'' denotes the Schwarzschild reference orbit. At fixed orbital elements, the radial and polar phase equations, Eqs.~\eqref{domega} and \eqref{dchi}, are independent of Finslerian parameter $\epsilon$, whereas the azimuthal equation \eqref{dvarphi2} contains an additional term. Consequently,
\begin{equation}
	\omega - \omega_{\rm S}	=0	,	\qquad	\chi - \chi_{\rm S} =0	,	\qquad	\varphi - \varphi_{\rm S} = \epsilon \psi(t)	,
\end{equation}
where
\begin{equation}
	\psi(t) = \int^t_{t_0} \Omega(t') dt' = \chi(t) - \chi_0	,	\qquad	\Omega \equiv \frac{d\chi}{dt}	.
\end{equation}
The trajectories are therefore related by a time dependent rotation about the symmetry axis,
\begin{equation}\label{transforms}
	\boldsymbol{x} = \mathsf{R} \boldsymbol{x}_{\rm S}, \qquad \mathsf{R} \equiv \begin{pmatrix}
 \cos(\epsilon\psi)&-\sin(\epsilon\psi)&0\\
 \sin(\epsilon\psi)& \cos(\epsilon\psi)&0\\
 0&0&1
 \end{pmatrix}	.
\end{equation}
The orthogonality relation $\mathsf{R}^{\rm T} \mathsf{R} = \mathsf{1}_{3}$ implies $r = r_{\rm S}$, where $\mathsf{1}_{3}$ is the $3 \times 3$ identity matrix. Differentiating the position relation gives
\begin{equation}
	\boldsymbol{v} = \mathsf{R} \left( \boldsymbol{v}_{\rm S} + \epsilon \boldsymbol{v}_1 \right),	\qquad	\mathsf{W} = \begin{pmatrix}
 0& -1& 0\\
 1&  0& 0\\
 0&  0& 0
 \end{pmatrix}	,
\end{equation}
where $ \boldsymbol{v}_1 = \Omega \mathsf{W} \boldsymbol{x}_{\rm S}$ arises from the time dependence of the rotation matrix, with $\dot{\mathsf{R}} = \epsilon \Omega \mathsf{R} \mathsf{W}$.
Using the position transformation $\boldsymbol{x} = \mathsf{R} \boldsymbol{x}_{\rm S}$ and the orthogonality of $\mathsf{R}$, the symmteic trace-free mass quadrupole can be written as
\begin{equation}
	\mathsf{Q} = \mu \left( \boldsymbol{x} \boldsymbol{x}^{\rm T} - \frac{r^2}{3} \mathsf{1}_{3} \right)= \mathsf{R} \mathsf{Q}_{\rm S} \mathsf{R}^{\rm T}	,
\end{equation}
To account for the time dependence of the rotation in successive derivatives, we define $\mathsf{Q}^{(n)} \equiv \mathsf{R} \mathsf{Y}_{n} \mathsf{R}^{\rm T}$, with $\mathsf{Y}_0 = \mathsf{Q}_{\rm S}$. Differentiation yields the recurrence relation
\begin{equation}
	\mathsf{Y}_{n+1} = \dot{\mathsf{Y}}_n + \epsilon \Omega \left[ \mathsf{W}, \mathsf{Y}_n \right]	,	\qquad	\left[ \mathsf{W}, \mathsf{Y}_n \right] = \mathsf{W}  \mathsf{Y}_n - \mathsf{Y}_n \mathsf{W}	.
\end{equation}
Here and below, overdots and superscripts in parentheses denote derivatives with respect to coordinate time. Keeping $\mathsf{R}$ unexpanded and retaining the remaining terms through first order in $\epsilon$, the quadrupole derivatives are then
\begin{equation}
	\mathsf{Q}^{(n)} = \mathsf{R} \left( \mathsf{Q}^{(n)}_{\rm S} + \epsilon \mathsf{C}_n \right) \mathsf{R}^{\rm T}	,
\end{equation}
with
\begin{equation}
	\mathsf{C}_n = \sum_{k=1}^{n} \begin{pmatrix} 
									n\\
									k
	\end{pmatrix} \Omega^{(k-1)} \left[ \mathsf{W}, \mathsf{Q}^{(n-k)}_{\rm S} \right]	.
\end{equation}

Using the fifth time derivative of the mass quadrupole, the Burke-Thorne generalized force takes the form~\cite{Thorne:1969rba,Burke:1970wx,Galley2009,Galley2012} 
\begin{equation}
	{\mathcal{F}}^{\rm RR} = -\frac{2\mu}{5} \mathsf{Q}^{(5)} \boldsymbol{x} = -\frac{2\mu}{5} \mathsf{R} \left( \mathsf{Q}^{(5)}_{\rm S} + \epsilon \mathsf{C}_5 \right) \boldsymbol{x}_{\rm S}	.
\end{equation}
To convert this force into a coordinate acceleration, we use the effective Lagrangian in auxiliary Cartesian coordinates,
\begin{equation}
	\mathcal{L} = \frac{\mu}{2} \boldsymbol{v}^2 + \frac{\mu M}{r} -\mu \epsilon \frac{h h_z}{r^2}	,
\end{equation}
where $h = |\boldsymbol{h}|$, $\boldsymbol{h} = \boldsymbol{x} \times \boldsymbol{v}$, and $h_z = \boldsymbol{e}_z \cdot \boldsymbol{h}$. Then, the forced Euler-Lagrange equation can be written as
\begin{equation}
	\mathsf{\mathcal{H}}\boldsymbol{a} + \boldsymbol{\mathcal{C}}(\boldsymbol{x},\boldsymbol{v}) = {\mathcal{F}}^{\rm RR}	,
\end{equation}
where 
\begin{equation}
	\mathsf{\mathcal{H}} \equiv \frac{\partial^2\mathcal L} {\partial\boldsymbol{v} \partial\boldsymbol{v}^{\rm T}} = \mu \left( \mathsf{1}_3 - \epsilon \mathsf{\mathcal{H}}_1 \right)	,	\qquad	\mathsf{\mathcal{H}}_1 \equiv \frac{\partial^2\mathcal{}} {\partial\boldsymbol{v} \partial\boldsymbol{v}^{\rm T}} \left(\frac{h h_z}{r^2}\right)	,
\end{equation}
and 
\begin{equation}
	\boldsymbol{\mathcal{C}} \equiv \frac{\partial(\nabla_{\boldsymbol{v}} \mathcal{L})}{\partial\boldsymbol{x}^{\rm T}} \boldsymbol{v} -\nabla_{\boldsymbol{x}} \mathcal{L}.
\end{equation}
The partial derivatives are taken with position and velocity treated as independent variables. The acceleration can be decomposed into $\boldsymbol{a} = \boldsymbol{a}^{\rm cons} + \boldsymbol{a}^{\rm RR}$. Subtracting the conservative equation, $\mathsf{\mathcal{H}} \boldsymbol{a}^{\rm cons} + \boldsymbol{\mathcal{C}} = 0$, therefore gives
\begin{equation}
	\boldsymbol{a}^{\rm RR} = \mathsf{\mathcal{H}}^{-1} \mathcal{F}^{\rm RR}	.
\end{equation}
Under simultaneous rotational transformation of position and velocity about the symmetry axis, its velocity Hessian transforms by conjugation with $\mathsf{R}$. Accounting for the additional $\mathcal{O}(\epsilon)$ term in the velocity transformation, we obtain
\begin{equation}
	\mathsf{\mathcal{H}}_1 = \mathsf{R} \left( \mathsf{\mathcal{H}}_{1, \rm S} + \mathcal{O}(\epsilon) \right) \mathsf{R}^{\rm T}	,	\qquad	\mathsf{\mathcal{H}}_{1,\rm S} \equiv \frac{\partial^2\mathcal{}} {\partial\boldsymbol{v}_{\rm S} \partial\boldsymbol{v}^{\rm T}_{\rm S}} \left(\frac{h_{\rm S} h_{z, \rm S}}{r_{\rm S}^2}\right)	.
\end{equation}
Since $\mathsf{\mathcal{H}}_1$ enters $\mathsf{\mathcal{H}}$ multiplied by $\epsilon$, evaluating this correction on the Schwarzschild reference orbit is sufficient at the retained order. Thus,
\begin{equation}
	\mathsf{\mathcal{H}}^{-1} = \frac{1}{\mu} \mathsf{R} \left( \mathsf{1}_3 + \epsilon \mathsf{\mathcal{H}}_{1, \rm S} \right) \mathsf{R}^{\rm T}	.
\end{equation}
Combining this expression with the generalized force yields
\begin{equation}\label{a^RR}
	\boldsymbol{a}^{\rm RR} = -\frac{2}{5} \mathsf{R} \left[\mathsf{Q}^{(5)}_{\rm S} + \epsilon \left( \mathsf{C}_5 +  \mathsf{\mathcal{H}}_{1,\rm S} \mathsf{Q}^{(5)}_{\rm S} \right) \right] \boldsymbol{x}_{\rm S}	.
\end{equation}
Thus, the radiation reaction acceleration in the Finslerian spacetime can be expressed in terms of Schwarzschild reference quantities and the rotation matrix $\mathsf{R}$.

We retain only the leading order Finslerian corrections and neglect mixed Finslerian-PN contributions. Since the Finslerian correction enters at Newtonian order, it suffices to evaluate the reference quantities on a Newtonian elliptical orbit. To simplify the calculation, we choose the $x$-axis along the ascending node and introduce the orthonormal orbital basis
\begin{align}
	\boldsymbol{n} &= \left( \cos\chi, \cos\iota \sin\chi, \sin\iota \sin\chi \right)	,	\\
	\boldsymbol{t} &= \left( -\sin\chi, \cos\iota\cos\chi, \sin\iota\cos\chi \right)	,	\\
	\boldsymbol{\ell} &= \left(0, -\sin\iota, \cos\iota\right)	,
\end{align}
where $\boldsymbol{n}$, $\boldsymbol{t}$, and $\boldsymbol{\ell}$ denote the radial, transverse, and orbital-normal unit vectors, respectively. Defining $u = 1 + e\cos\omega$ and $\rho = e \sin\omega$, the reference position, velocity, and polar frequency are
\begin{equation}
	\boldsymbol{x}_{\rm S} = \frac{M p}{u} \boldsymbol{n}	,	\qquad	\boldsymbol{v}_{\rm S} = \frac{\rho \boldsymbol{n} + u \boldsymbol{t}}{\sqrt{p}}	,	\qquad	\Omega = \frac{u^2}{M p^{3/2}}	.
\end{equation}
The three contributions to Eq.~\eqref{a^RR} are then
\begin{align}
	\mathsf{Q}^{(5)}_{\rm S} \boldsymbol{x}_{\rm S} =& -\frac{8 \mu u^3 \rho}{3M^2 p^{9/2}}\left(3 \rho^2 + 18 u^2 + 4 u\right) \boldsymbol{n} + \frac{4 \mu u^4}{M^2 p^{9/2}} \left( 6 u^2 - 2 u - 9 \rho^2 \right) \boldsymbol{t}	,	\\
	\mathsf{C}_5 \boldsymbol{x}_{\rm S} =& \frac{4 \mu u^4}{M^2 p^{9/2}} \left[ -10 u \rho \cos\iota \left( 4 u + 3 \right) \boldsymbol{n} + \cos\iota \left( 16 u^3 + 6 u^2 - 2 u - 40 u \rho^2 - 9 \rho^2 \right) \boldsymbol{t}	\right.\nonumber	\\
	&\left. + \sin\iota \left(5 u \rho \sin\chi \left( 4 u + 3 \right) - \cos\chi \left( 6 u^3 + 6 u^2 - 2 u - 10 u \rho^2 - 9 \rho^2 \right) \right) \boldsymbol{\ell} \right]	,	\\
	\mathsf{\mathcal{H}}_{1,\rm S} \mathsf{Q}^{(5)}_{\rm S} \boldsymbol{x}_{\rm S} =& \frac{4 \mu u^4}{M^2 p^{9/2}} \left( 6 u^2 - 2 u - 9 \rho^2 \right)\left( 2 \cos\iota \, \boldsymbol{t} - \sin\iota \cos\chi \, \boldsymbol{\ell} \right)	.
\end{align}
For subsequent calculations, we write $\boldsymbol{a}_{\rm S} = -\frac{2}{5}\mathsf{Q}^{(5)}_{\rm S} \boldsymbol{x}_{\rm S}$ and $\boldsymbol{a}_1 = -\frac{2}{5} \left( \mathsf{C}_5 + \mathsf{\mathcal{H}}_{1,\rm S} \mathsf{Q}^{(5)}_{\rm S} \right)  \boldsymbol{x}_{\rm S}$.

We now evaluate the radiation reaction contributions to Eqs.~\eqref{eff_dE}, \eqref{eff_dJ}, and \eqref{eff_dK} explicitly. In the terms on the right-hand sides that are explicitly proportional to $\epsilon$, the relevant quantities can be replaced by their general relativity counterparts. The first term on the right-hand side of each equation is evaluated in detail below. For the energy equation \eqref{eff_dE},
\begin{align}
	\boldsymbol{v} \cdot \boldsymbol{a}^{\rm RR} = \left[ \mathsf{R} \left(\boldsymbol{v}_{\rm S} + \epsilon \boldsymbol{v}_1 \right) \right] \cdot \left[ \mathsf{R} \left(\boldsymbol{a}_{\rm S} + \epsilon \boldsymbol{a}_1 \right) \right] = \boldsymbol{v}_{\rm S} \cdot \boldsymbol{a}_{\rm S} + \epsilon \left( \boldsymbol{v}_{\rm S} \cdot \boldsymbol{a}_{\rm 1} + \boldsymbol{v}_{\rm 1} \cdot \boldsymbol{a}_{\rm S} \right) + \mathcal{O}(\epsilon^2)	.
\end{align}
Using the orthonormal orbital basis, we obtain
\begin{align}
	\boldsymbol{v}_{\rm S} \cdot \boldsymbol{a}_{\rm S} &=	-\frac{8 \mu u^3}{15 M^2 p^5} \left( 18 u^4 - 6 u^3 - 8 u \rho^2 - 63 u^2 \rho^2 - 6 \rho^4 \right)	,	\\
	\boldsymbol{v}_{\rm S} \cdot \boldsymbol{a}_{\rm 1} + \boldsymbol{v}_{\rm 1} \cdot \boldsymbol{a}_{\rm S} &= -\frac{16 \mu u^5 \cos\iota}{5 M^2 p^5} \left(12 u^2 + 8 u^3 - 33 \rho^2 - 4 u - 40 u \rho^2\right)	.
\end{align}
Using the position and velocity transformations, we obtain dissipative rates of the angular momentum magnitude and its axial component,
\begin{equation}
	\dot{h} = \frac{\boldsymbol{h}}{h} \cdot \left( \boldsymbol{x} \times \boldsymbol{a}^{\rm RR} \right)	,	\qquad	\dot{h}_z = \boldsymbol{e}_z \cdot \left( \boldsymbol{x} \times \boldsymbol{a}^{\rm RR} \right)	.
\end{equation}
By making use of orthonormal orbital basis, we obtain
\begin{align}
	\dot{h} =& -\frac{8 \mu u^3}{5 M p^{7/2}} \left( 6 u^2 - 2 u - 9 \rho^2 \right) - \frac{8 \mu \epsilon u^3 \cos\iota}{5 M p^{7/2}} \left( 18 u^2 + 16 u^3 - 27 \rho^2 - 6 u - 40 u \rho^2 \right)	,	\\
	\dot{h}_z =& -\frac{8 \mu u^3 \cos\iota}{5 M p^{7/2}} \left( 6 u^2 - 2 u - 9 \rho^2 \right) - \frac{8 \mu \epsilon u^3}{5 M p^{7/2}} \left[ \cos^2\iota \left( 18 u^2 + 16 u^3 - 27 \rho^2 - 6 u - 40 u \rho^2 \right) \right. \nonumber \\
	&\left. + 2 \sin^2\iota \cos^2\chi \left( 6 u^2 + 3 u^3 - 9 \rho ^2 - 2 u - 5 u \rho^2\right) - 5 u \rho \sin^2\iota \sin\chi \cos\chi \left( 3 + 4 u \right) \right]	.
\end{align}
Combining these results with the remaining terms in Eqs.~\eqref{eff_dE}-\eqref{eff_dK}, we obtain the
\begin{align}
	\dot{E} =& -\frac{8 \mu^2 u^3}{15 M^2 p^5} \left( 18 u^4 - 6 u^3 - 8 u \rho^2 - 63 u^2 \rho^2 - 6 \rho^4 \right) \nonumber \\
	&- \frac{32 \mu^2 \epsilon u^5 \cos\iota}{5 M^2 p^5} \left( 4 u^3 + 3 u^2 - u - 20 u \rho^2 - 12 \rho^2 \right)	,	\label{eff_dE2}	\\
	\dot{J} =& -\frac{8 \mu^2 u^3 \cos\iota}{5 M p^{7/2}}\left( 6 u^2 - 2 u - 9 \rho^2 \right) - \frac{8 \mu^2 \epsilon u^3}{5 M p^{7/2}} \left[ \cos^2\iota \left( 16 u^3 + 6 u^2 - 2 u - 40 u \rho^2 - 9 \rho^2 \right) \right. \nonumber \\
	&\left. + \sin^2\iota \cos^2\chi \left( 6 u^3 + 6 u^2 - 2 u - 10 u \rho^2 - 9 \rho^2 \right) - 5 u \rho \sin^2\iota \sin\chi \cos\chi \left( 3 + 4 u \right) \right]	,		\label{eff_dJ2}	\\
	\dot{K} =& -\frac{8 \mu^2 u^3}{5 M p^{7/2}}\left( 6 u^2 - 2 u - 9 \rho^2 \right) - \frac{16 \mu^2 \epsilon u^3 \cos\iota}{5 M p^{7/2}} \left(6 u^2 + 8 u^3 - 9 \rho^2 - 2 u - 20 u \rho^2\right)	. 	\label{eff_dK2}	
\end{align}

\bibliography{reference}
\end{document}